\documentclass[fleqn,twoside]{article}
\usepackage{espcrc2}

\title{Phase contribution to LFV\thanks{Presentation given by K.Tsumura}}

\author{S. Kanemura\address[osaka]{Department of Physics,
		Osaka University Toyonaka, Osaka 560-0043, Japan},
        K. Matsuda\addressmark[osaka],
        T. Ota\addressmark[osaka],
        T. Shindou\address{Theory group, KEK, Tsukuba 305-0801, Japan},
        E. Takasugi\addressmark[osaka]
        and K. Tsumura\addressmark[osaka]\thanks{ko2@het.phys.sci.osaka-u.ac.jp}}
           
\begin{document}
\maketitle
Lepton flavor violation (LFV) is an inportant sign of new physics.
Especially, some supersymmetric models naturally predict sizable LFV observables.  
 
We consider the minimal supersymmetric standard model (MSSM) with right-handed neutrinos.
The soft supersymetry breaking terms are assumed to be universal at the GUT scale $M_{GUT}$.
In this framework, we can parameterize the neutrino Yukawa matrix $Y_\nu$ in terms of physical quantities as
\begin{eqnarray}
\frac{v}{\sqrt{2}} Y_\nu = \sqrt{M_R^{diag}} R \sqrt{m_\nu^{diag}} U_{MNS}^\dag ,
\end{eqnarray}
where $v$ is the vacuum expectation value of the Higgs boson, $U_{MNS}$ is the Maki-Nakagawa-Sakata (MNS) matrix including two Majorana phases,
$m_\nu^{diag}$ is neutrino mass matrix and $M_R^{diag}$ is right-handed neutrino mass matrix in each diagonal basis.
Arbitrary complex orthogonal matrix $R$ is expressed as
$R \equiv O_{12} O_{23} O_{31} Q_{12} Q_{23} Q_{31}$, where
\begin{eqnarray} 
O_{12}(\theta_{12})\equiv \left( \begin{array}{ccc} \cos \theta_{12} & - \sin \theta_{12} & \\
                                  \sin \theta_{12} & \cos \theta_{12} & \\
                                    & & 1 \end{array} \right), \rm{etc.},
\end{eqnarray}
and $Q_{ij} \equiv O_{ij}(i \eta_{ij})$.
The elements $Q_{ij}$ strongly affect the Yukawa matrix as hyperbolic functions.

In the following, we assume that the neutrino masses are approximately degengerate and the right-handed neutrino masses are also degenerate $(m_\nu^{diag} \sim m \cdot 1,M_R^{diag} \sim M \cdot 1)$.
We then obtain $Y_\nu$ in the diagonal form as
\begin{eqnarray}
Y_\nu^{diag} \sim \frac{\sqrt{2}}{v} \sqrt{M m}
\left(
\begin{array}{ccc}
1/r & & \\ & 1 & \\ & & r
\end{array}
\right),
\end{eqnarray}
where,
$r^2 \equiv 2x^2-1+2x \sqrt{x^2-1}$, $x \equiv c_{12}c_{23}c_{31}$, and $c_{ij} \equiv \cosh \eta_{ij}$.
For $m \sim 0.1 \rm{eV}$ and $M \sim 10^9 \rm{GeV}$, $r$ is close to ${\cal O}(10^2)$, so that
the neutrino Dirac mass (Yukawa coupling) spectrum becomes hierarchical.
It is similar to the up-type quark Yukawa hierarchy.
The parameter $r$ ,which appears due to the non-zero phases of $R$, determines the scale of the neutrino Yukawa matrix and Yukawa hierarchy.

The LFV processes $l_i \to l_j \gamma$ appear in the MSSM due to the slepton mixing $(\Gamma \propto |(m^2_{\tilde{L}})_{ij}|^2)$ 
which comes from the renormalization group effect on $Y_\nu$ between $M_{GUT}$ and $M_R$ .
The off-diagonal elements of the slepton mass matrix represent by $(m^2_{\tilde{L}})_{ij} \propto (Y_\nu^\dag Y_\nu)_{ij} (i \ne j)$.
The LFV branching ratio is proportional to $r^4$, so that a large value of $r$ can change the branching ratios by a few order of magnitude.
On the other hand, we numerically find that Majorana phases make the magnitude of the LFV branching ratios periodic, and its effect is not larger than that of $r$.

\end{document}